\documentclass[%
showkeys,
showpacs,
reprint,
superscriptaddress,
nofootinbib,
nobibnotes,
 amsmath,amssymb,
aps,
]{revtex4-2}
\usepackage{amssymb}
\usepackage{color}
\usepackage{graphicx}
\usepackage{dcolumn}
\usepackage{bm}
\usepackage{hyperref}
\usepackage{breakurl}

\begin{document}

\title{Experimental measurement of infinite dilution thermal neutron self-shielding factor}

\author{Ateia~W.~Mahmoud}\email{atia.mahmoud@gmail.com} \email{ateia.mahmoud@eaea.org.eg}
\address{Physics Department, Faculty of Science, Ain Shams University, Cairo, Egypt.}
\address{Reactor Physics Department, Nuclear Research Center, Egyptian Atomic Energy Authority, Cairo 13759, Egypt.}

\author{Elsayed~K.~Elmaghraby}\email[Corresponding Author: ]{e.m.k.elmaghraby@gmail.com}
\email{elsayed.elmaghraby@eaea.org.eg}
\address{Experimental Nuclear Physics Department, Nuclear Research Center, Egyptian Atomic Energy Authority, Cairo 13759, Egypt.}

\author{E.~Salama}\email{e\_elsayed@sci.asu.edu.eg}
\address{Basic Science Department, Faculty of Engineering, The British University in Egypt (BUE), Cairo, Egypt.}

\author{A.~Elghazaly}\email{an\_4558@yahoo.com}
\address{Reactor Physics Department, Nuclear Research Center, Egyptian Atomic Energy Authority, Cairo 13759, Egypt.}

\author{S.~A.~El-fiki}\email{soadelfiki@sci.asu.edu.eg}
\address{Physics Department, Faculty of Science, Ain Shams University, Cairo, Egypt.}

\date{\today}

\begin{abstract}
The absorption of neutrons in media together with its transport properties cause the neutron flux to decrease as it penetrates the material because the absorption of neutrons in the sample itself attenuates the neutrons flux as it goes deeper into the sample. In the present work, the thermal neutron self-shielding factor of indium, gold, zinc, and mercury were determined experimentally. The current results together with those found in the literature were used to validate a mathematical \emph{ab initio} formulae based on integral cross-section parameters  used to compare our results. The complete agreement among these species of data suggests the validity of correlating the neutron migration length in the convex-shaped material with the average chord length described in the mathematical model.
\end{abstract}


\maketitle


\section{Introduction}
Neutron activation analysis has evolved into a highly effective nuclear analytical technique over time. However, the minimum observable activity by neutron activation within a valid measurement interval is constrained by a number of factors such as neutron fluence, the fraction of fluence that reaches the interior of the sample, and sample geometry \cite{TohamyElmaghrabyComsan2020109340,NAKAMURA2013119,AliandElmaghraby202063,FarinaArbocco2012,Elmaghraby2018PhysScr,Chilian2010429,Elmaghraby2019PhysScrCode,Elmaghraby2016Shape,Jacimovic2010399,Elmaghraby201742} .Non-destructive elemental concentration measurement in unknown materials and nuclear material interrogation are common applications for such techniques \cite{TohamyElmaghrabyComsan2019162387,TohamyElmaghrabyComsan2021045304}. when exposing a sample to a neutron beam, the interior of the sample will be exposed to a lesser neutron fluence rate than the exterior part, in all circumstances regardless of the geometry of the neutron current. This phenomenally known as \emph{self-shielding}, and it is a critical element of the neutron transport phenomena. In the case of field geometry, the net neutron current essentially disappears, while the fluence (or flux) becomes the observable quantity.  There is an interplay between neutron absorption in the sample and the overall neutron flux \cite{Elmaghraby2021ICPAP2021C1}.  Predominantly, the correlation between neutron self-shield factors and the set of parameters involved in the calculation of its value had been studied by several scientists\cite{Fleming19821263,Blaauw1995403B,Gonalves2001447,Martinho2003371,Salgado2004426,Martinho2004637,Goncalves2004186,Sudarshan2005205,Nasrabadi2007473,Mashkovich1983Book,Moll2020106990,MahmoudElmaghrabySalamaElghazalyElFikiarxiv220413239}, who gave dimensionless variables to identify and encompass the physical and geometric varieties of the sample's geometries to attain a universal formula for self-shielding. In this study, neutron self-shielding factors for Indium and gold samples were determined experimentally.

\section{Materials and Methods}

\subsection{Samples}
Indium metal in the form of sheets (purity 99.9+\% purchased from Aldrich was used. Some foils were processed to reach thicknesses of 0.001 cm by rolling them repeatedly until reaching the desired thickness. The 0.001 cm foils were used as infinite diluted reference samples. The absorber samples were indium sheets each of thicknesses 0.0175 cm for which the self-shielding factor shall be calculated.  Gold foils (purity of 99.999\% purchased from  Aldrich and having thicknesses of 0.002 cm were used as infinite diluted samples. While stacked sheets of these foils having an overall thickness of 0.01 cm were used as the self-shielded sample.  The infinite diluted samples were used as neutron monitors as well. For zinc, a metal foil of a thickness of 0.066 cm was used as a reference sample for infinite dilution. The absorber sample of zinc was a cylinder having a radius of 1 cm and a height of 1 cm. Also, cylinders of mercury, one with a radius of 0.3 cm and a height of 0.27 cm, were used as a reference for infinite dilution and another with a radius of 0.3 cm and a height of 3.89 cm as the absorbent sample.

The investigated activation reactions are $^{115}$In(n,$\gamma$)$^{116m}$In ($T_{1/2}$= 54.29 min, $E_\gamma$=1293.6 keV, $I_\gamma$=84.3\% \cite{Blachot2010717}, $\sigma_{a}$= 202.2 b \cite{Elmaghraby2018PhysScr}, $\sigma_g^{(m)}$=162.6 b \cite{Elmaghraby2018PhysScr}, $I_g$=2585 b \cite{Elmaghraby2016Shape,Elmaghraby2019PhysScrCode}),  $^{197}$Au(n,$\gamma$)$^{198}$Au ($T_{1/2}$= 2.69 d, $E_\gamma$=411.8 keV, $I_\gamma$=95.58\% \cite{Huang2016221}, $\sigma_{a}$= $\sigma_g$=98.65 b, $I_g$=1567 b \cite{Elmaghraby2016Shape,Elmaghraby2019PhysScrCode}),
$^{68}$Zn(n,$\gamma$)$^{69}$Zn ($T_{1/2}$=  13.8 h, $E_\gamma$=438.634 keV, $I_\gamma$=94.849\% \cite{Nesaraja20141}, $\sigma_{a}$= $\sigma_g$= 1.065 b, $I_g$= 3.063 b \cite{Elmaghraby2016Shape,Elmaghraby2019PhysScrCode}), and
$^{202}$Hg(n,$\gamma$)$^{203}$Hg ($T_{1/2}$=  46.594 d, $E_\gamma$= 279.1952 keV, $I_\gamma$=81.56\% \cite{Kondev20051}, $\sigma_{a}$= $\sigma_g$= 4.903 b, $I_g$= 33.298 b \cite{Elmaghraby2016Shape,Elmaghraby2019PhysScrCode}). Here, $T_{1/2}$ is the half-life of the activation nucleus, $E_\gamma$ is its the specific gamma-line, $I_\gamma$ is its intensity, $\sigma_g$ is the capture cross-section, $\sigma_{a}$ is the absorption cross-section, and $I_g$ is the resonance integral.

\subsection{Neutron irradiation}
There is a common difficulty to find an isotropic neutron source with constant fluence rate. For this purpose, we used previously made setup of two AmBe neutron sources embedded within moderating medium \cite{Elmaghraby2018PhysScr}. The two AmBe sources were placed in a cylindrical-shaped paraffin wax of diameter 58 cm and height of 57 cm (as moderator) along the moderator diameters and parallel to its axis at two positions separated by a distance of 16 cm apart (effective center to center distance is 20.4 cm). {The two sources are identical in activities and deliver axially symmetric neutron fluence at the irradiation tube. The alpha-beryllium neutron production physics and the long half-life of $^{241}$Am guarantee a constant fluence rate for the experiment. The irradiation time of each sample was 5 times its half-life }.

\subsection{Gamma-ray measurements}
The gamma-ray count rates were measured using ORTEC 70\%  HPGe detector; energy resolution was about 4.1 keV at the 1332.5 keV $\gamma$-line of $^{60}$Co. {The data acquisition and analysis were performed using Genie 2000 Spectroscopy Software. Acquisition times were between 30 min and 1 hour, depending on the decay constants of the activation isotopes. The detector calibration was performed using a set of standard point sources of $^{60}$Co, $^{137}$Cs, $^{133}$Ba, $^{241}$Am, $^{22}$ Na (OXFORD instrument Inc.)}. Lead shield having an inner concentric copper cylinder of thickness 6 mm was used to reduce the $x$-rays originated from gamma interaction with the lead material, lined 3 mm aluminum cylinder to ensure maximum reduction of x-ray and consequently reduction of background.

\subsection{Uncertainty}
\label{Sec:uncertainty}
The systematic uncertainty in our measurements lies in the determination of the neutron flux which would be affected by the self-shielding of the monitor itself. In order to reduce such a shift, we used the thin indium and gold foils as our reference. The statistical uncertainties in gamma-ray measurements were lower than 2\% and dead time for all measurements was less than 0.1\%. Uncertainties in the reference standard activities were below 5\%.  The uncertainty in mass measurements was 0.0005 g. The statistical uncertainties associated with each of our measurements were added in quadratic error propagation form in accordance with the \emph{Provisional Rule}~\cite{Taylor1997BookStat}. The total experimental uncertainty was below 10\%.

\section{Results and discussion}
\subsection{Measurement of self-shielding factor}
The thermal neutron self-shielding factor,  abbreviated $G_{\mathrm{th}}$, is the ratio of the volume-averaged fluence rate within the material's volume that may absorb or scatter neutrons to the fluence rate within the same volume in the absence of neutron interaction\cite{Blaauw1995403B,Gonalves2001447,Martinho2003371,Martinho2004637,Goncalves2004186}. As a result, for the definition, the researcher utilizes the reaction rate per atom rather than the fluence rate. Thermal neutron capture reactions dominate the higher energy reaction processes when irradiating a single material without cadmium cover, implying that the reaction rate is primarily due to thermal neutrons. In this situation, the flux detected by the sample is the flux normalized by the thermal neutron self-shielding factor i.e. $\varphi_t$=$G_{th}\varphi_o$ is the self-shielded flux of neutrons for samples having thickness t. (For ultrathin foils $G_{th}\sim 1$.)

Based on the properties of the dimensional quantity of the average chord length ($\bar{\ell}$), which has the greatest impact on the medium's self-shielding proprieties \cite{MahmoudElmaghrabySalamaElghazalyElFikiarxiv220413239}, the values of thermal neutron self-shielding factors for thick sheets of indium, gold, zinc, and cylinders of mercury were determined experimentally relative to the thin foils made of the same materials. The reaction rates per atom ($RR(\bar{\ell})=\phi_t\sigma_g$) were determined after the neutron irradiation of each sample, individually, using the following relations:
\begin{equation}\label{Eq:tares0}
[G_{th}(\bar{\ell})]_{\text{infinite dilution}}=\frac{{RR}(\bar{\ell}_{\text{large}})}{{RR}(\bar{\ell}_{\text{small}})},
\end{equation}
\noindent where
\begin{equation}\label{Eq:tares1}
{RR}(\bar{\ell}_{\text{large}})=\frac{M {{C}_{\gamma}}{{\lambda}_{\gamma}}}{\varepsilon_\gamma I_\gamma N_A f m (1\text{-}e^{\text{-}\lambda_\gamma t_{i}})e^{\text{-}\lambda_\gamma t_{c}}(1\text{-}e^{\text{-}\lambda_\gamma t_{m}})}
\end{equation}
is the reaction rate with the parameters specified for sample having larger dimension, and
\begin{equation}\label{Eq:tares2}
{RR}(\bar{\ell}_{\text{small}})=\frac{M {{C}_{\gamma}}{{\lambda}_{\gamma}}}{\varepsilon_\gamma I_\gamma N_A f m (1\text{-}e^{\text{-}\lambda_\gamma t_{i}})e^{\text{-}\lambda_\gamma t_{c}}(1\text{-}e^{\text{-}\lambda_\gamma t_{m}})}
\end{equation}
is the reaction rate with the parameters specified for  sample having smallest possible dimension. Here $\lambda_\gamma$ represents the decay constant of the residual nucleus that emits the specific $\gamma$-line,  $\varepsilon_\gamma$ is the detector's efficiency,and $C_{\gamma}$ is the photo-peak area measured during the time $t_{m}$.

The irradiation time ($t_{i}$) and cooling time ($t_{c}$) are adopted according to the half-life of the residual isotope.  The number of atoms in the sample is determined using the relation  $N_{a}=N_{A} f m / M$, where $N_{A}=$$6.02 \times 10^{23} atom/mol$
is the Avogadro's number, $M$ is the molar mass of the target molecules, $f$ is the isotopic abundance of the target atom ($f$=1 in the case of Au while $f$=0.9572 for $^{115}$In). The experimental conditions of the present experiments are given in Table \ref{Tab:ExperimentalResults}.

{
\begin{table}
  \centering
  \caption{Experimental conditions for thermal neutron self-shielding factors of indium and gold samples.}\label{Tab:ExperimentalResults}
\begin{tabular}{rrr}
\hline\hline
  Absorber &    dimensions  &      $\bar{\ell}$ (cm)  \\
\hline
$^{115}$In(n,$\gamma$)$^{116m}$In &    t=0.001 cm &             0.003   \\
           &    t=0.0175 cm &          0.0525 \\
infinite dilution chord-length &     &             0.0495   \\
\hline
$^{197}$Au(n,$\gamma$)$^{198}$Au &    t=0.002 cm &           0.006   \\
           &       t= 0.01 cm &             0.03  \\
infinite dilution chord-length &     &             0.024   \\
\hline
$^{68}$Zn(n,$\gamma$)$^{69}$Zn &  t= 0.066 cm    &  0.198  \\
           &  R=1 cm, H=1 cm     & 1.6993  \\
infinite dilution chord-length &     &   1.5013  \\
\hline
$^{202}$Hg(n,$\gamma$)$^{203}$Hg &  R=0.3 cm, H=0.27 cm    &  0.479 \\
           &   R=0.3 cm, H=3.89 cm     &  1.068  \\
infinite dilution chord-length &     &    0.5885   \\
\hline
\end{tabular}
\end{table}}

For indium foils, the average reaction rate per atom of the infinite diluted samples having $\bar{\ell}_{\text{small}}=$ 0.003 cm, as calculated by Eqs. \ref{Eq:tares1} and \ref{Eq:tares2}, is found to be 2.95$\pm$0.31$\times10^{-19}$ s$^{-1}$. On the other hand, the reaction rate per atom in the absorber sample having $\bar{\ell}_{\text{large}}=$ 0.0525 cm was found to be 2.55$\pm$0.081 $\times10^{-19}$ s$^{-1}$. The corresponding self shielding factor obtained from Eq. \ref{Eq:tares0} becomes 0.865$\pm$0.095. 

In the case of gold foils, the average reaction rate per atom of  the infinite diluted sample, having $\bar{\ell}_{\text{small}}=$ 0.006 cm was found to be 8.65$\pm$0.20$\times10^{-19}$ s$^{-1}$, while the reaction rate per atom in the absorber sample having $\bar{\ell}_{\text{large}}=$ 0.03 cm was found to be 8.08$\pm$0.08$\times10^{-19}$ s$^{-1}$. Hence, the self shielding factor obtained from Eq. \ref{Eq:tares0} becomes 0.934$\pm$0.023.

Experimental determinations of the self-shielding factors for zinc and mercury were difficult to  correlated to the infinite dilution samples due to the small cross-section of the monitored reactions $^{68}$Zn(n,$\gamma$)$^{69}$Zn and $^{202}$Hg(n,$\gamma$)$^{203}$Hg, respectively, as given in Table \ref{ElementCrossSections}. Hence, the value of $\bar{\ell}_{\text{small}}$ reference sample may be forced to be large to reach measurable values of the activities of the monitored reactions.
In the present experiments, the average reaction rate per atom for zinc foils having $\bar{\ell}_{\text{small}}=$ 0.198 cm was found to be 0.0124$\pm$0.001$\times10^{-19}$ s$^{-1}$, while the reaction rate per atom of the absorber zinc cylinder sample having $\bar{\ell}_{\text{large}}=$ 1.6993 cm was found to be 0.0109$\pm$0.001$\times10^{-19}$ s$^{-1}$.
Similarly, the average reaction rate per atom of reference mercury cylinders having $\bar{\ell}_{\text{small}}=$ 0.479 cm was determined to be 0.103$\pm$0.021$\times10^{-19}$ s$^{-1}$ while the reaction rate per atom of the absorber sample having $\bar{\ell}_{\text{large}}=$ 1.068 cm was found to be 0.0195$\pm$0.001$\times10^{-19}$ s$^{-1}$. The self shielding factor calculated from Eq. \ref{Eq:tares0} is 0.881$\pm$0.115 for zinc samples, while for mercury samples, it is 0.1898$\pm$0.0397.

\subsection{Comparison with mathematical model}
To make a comparison with models, we had used our \emph{ab initio} \cite{MahmoudElmaghrabySalamaElghazalyElFikiarxiv220413239} in which we proposed that the thermal neutron self-shielding factor can be calculated from the following equation that have a sigmoid shape.
\begin{equation}\label{Eq:Selfshielding}
  G_{\mathrm{th}} =\left(\frac{\Sigma_t}{\Sigma_s}\right)\times\frac{1}{1+\mathcal{Z}} 
\end{equation}
\noindent  The factor $\left(\frac{\Sigma_t}{\Sigma_s}\right)$  represents the weight of the total interaction cross-section versus the scattering contribution. Microscopic values of cross-sections are given in Table \ref{ElementCrossSections}.
The dimensionless parameter, $\mathcal{Z}$, may be expressed as product of three functions, $ \mathcal{Z}=\Omega(\ldots)\chi(\ldots)\eta(\ldots) = (Geometry)(Composition)(Probability) $ ,the geometry function ($\Omega=\cfrac{\bar{\ell}}{P_0}$) is a function of the dimensions of the sample in the unit of [cm], macroscopic cross-section function ($\chi=\Sigma_t$) in the units of [cm$^{-1}$] which depends on the isotopic content of the sample, and a dimensionless neutron energy correcting factor ($\eta=\cfrac{\Sigma_a}{\Sigma_s}$) which is a function of the neutron absorption  and the scattering cross-sections.

\begin{table*}[htb]
	\caption{The reference cross-section and the element-averaged cross-section $\sigma_g$ and $\sigma_s$, are the capture and scattering cross-section at thermal energies,  $\sigma_a$,   $\sigma_t$  are the absorption and total cross-sections at thermal energies.}\label{ElementCrossSections}
	\centering
	\begin{small}
		\begin{tabular}{c|c c cccc}
			\hline\hline
			Element & Isotope & abundance & $\sigma_g$ &  $\sigma_s$ & $\sigma_a$ & $\sigma_t$ \\
			        &         & \% & (b)        & (b)         & (b)        & (b)        \\ \hline
			In      &   $^{115}$In & 95.72 &  162.6$^{(m)}$    &  2.519     &  202.2    &   204.72   \\
                    & &     &  39.6$^{(g)}$    &    &     &     \\
			      &   $^{113}$In & 4.28 &      &   3.683    &   12.13   &  15.81 \\
			     &   Element Average  &  &     & 2.569      & 194.07     & 196.64     \\
            \hline
			Au    & $^{197}$Au  & 100 & 98.672     & 7.9298      & 98.672     & 106.6      \\ \hline
			Zn    & $^{64}$Zn  & 49.2 &       & 3.927      & 0.787  & 4.71 \\
			      & $^{66}$Zn  & 27.7 &       & 4.938      & 0.618  & 5.56\\
			      & $^{67}$Zn  & 4    &       & 2.111      & 7.47   & 9.58\\
			      & \textbf{$^{68}$Zn} & 18.5 & 1.0572     & 4.2522 & 1.0572     & 5.31    \\
			      & $^{70}$Zn  &0.6   &       & 4.337      & 0.092  & 4.43\\
			      &  Element  Average  &      &       & 4.197           & 1.0533  & 5.25   \\ \hline
			Hg    & $^{196}$Hg & 0.15  &       & 109.8  & 3078   &  3187.8  \\
                  & $^{198}$Hg & 10.04 &       & 12.81  & 1.985  &  14.8    \\
                  & $^{199}$Hg & 16.94 &       & 66.7   & 2149   &  2215.7   \\
                  & $^{200}$Hg & 23.14 &       & 14.56  & 1.443  &  16  \\
                  & $^{201}$Hg & 13.17 &  & 13.8   & 4.903  &  18.7  \\
                  & \textbf{$^{202}$Hg} & 29.74&  4.954  & 14.61  & 4.954     &   19.56 \\
                  & $^{204}$Hg & 6.82 &     &   Unknown   &  0.85    & 0.85    \\
			      &  Element  Average  &  &     & 22.281     & 371.368  &393.65    \\
\hline
			
\multicolumn{6}{l}{$^{(m)}$ is the cross-section of formation of the metastable state $^{116m}$In}\\
\multicolumn{6}{l}{$^{(g)}$ is the cross-section of formation of the ground state $^{116g}$In}\\
		\end{tabular}
	\end{small}
\end{table*}

Here, $P_0$ is the probability of the first interaction derived from the transport kernel in steady-state \cite{Blaauw1996431,Stuart1957617} and averaged over the length the neutron passes in the material.
This path-length is the average neutron-chord length (denoted $\bar{\ell}$).

{The term $\Omega$  contains the total distance of interest within the medium, while $\chi$ and $\eta$ determine the slope of the steeping part of the curve. The macroscopic cross-section function is expressed as:
\begin{equation}\label{Eq:MacroscopicXSelemental}
 \chi(E_n) =\Sigma(E_1,E_2) = \frac{\rho N_A \theta_i}{M}  \tilde{\sigma} ,
\end{equation}

\noindent where $N_A$ is the Avogadro's number [mol$^{-1}$], $\rho$ is the density of the material [g cm$^{-3}$], and $M$ is its atomic mass [g mol$^{-1}$]. $\theta_i$ is the isotopic abundance of the absorbing isotope in which it should be multiplied  by the fraction of the element in the material if a compound material is used.  Here, $\tilde{\sigma}$ is the integral cross-section in the \emph{thermal energy} \cite{TohamyElmaghrabyComsan2020109340}}.

The $\eta$ term is the scattering to absorption ratio;
\begin{equation}\label{Eq:ScatteringTerm}
  \eta(\Sigma_s,\Sigma_a)=\frac{\Sigma_a}{\Sigma_s}
\end{equation}
{In general, values of thermal cross-section and resonance integral are well known from tables \cite{Elmaghraby2016Shape,Sukhoruchkin1998book,Sukhoruchkin2009NeutResPara} or integration of spectroscopic cross-section \cite{Elmaghraby2019PhysScrCode}}.

For approximation, the probability in the thermal energy range, $P_0$ becomes \cite{MahmoudElmaghrabySalamaElghazalyElFikiarxiv220413239}:
\begin{equation}\label{Eq:ApproximateP0}
P_0\sim(1-\exp(-\Sigma_t \bar{\ell})),
\end{equation}

Obtaining the orientation-dependent chord lengths of a convex body (which most materials have) is a complicated mathematical argument. However, most researchers treat the problem with the body that has the minimal volume in a class of convex bodies having the same dimensions \cite{Horvath20201}. We will use the simple formulation of the average neutronchord length based on the fact that the trajectories of the incident isotropic neutrons traverse different lengths within the body due to scattering are comparable \cite{MahmoudElmaghrabySalamaElghazalyElFikiarxiv220413239}, as long as we want to avoid the rigorous derivations of the average neutronchord length (cf. Refs. \cite{Mazzolo20031391,Roberts19994953,Dekruijf2003549,Zoia201920006,Khaldi201713,Zhang1999985} for details). Based on those results, the average neutron-chord length ($\bar{\ell}$) is taken for infinite foils and cylindrical shapes as follows:
\begin{equation}\label{Eq:ellinfinitesheet}
\bar{\ell}=3t,
\end{equation}

\begin{equation}\label{Eq:ellfinitcylinder-reduced}
  \bar{\ell}=\frac{3\pi R H}{2.4048 H +\pi R},
\end{equation}
where $t$ is the thickness of the foil, while $R$ and $H$ are the radius and height of the cylinder.
The cross-section used in the calculation of the model is illustrated in Table \ref{ElementCrossSections} based on earlier studies  \cite{Elmaghraby2016Shape,Elmaghraby2017Treat,Elmaghraby2018PhysScr,Elmaghraby2019PhysScrCode} specifically for the thermal neutron range.

One of the most important benefits of defining the chord length by Eqs. \ref{Eq:ellinfinitesheet} and \ref{Eq:ellfinitcylinder-reduced} is the possibility to use it to obtain an infinite-dilution self-shielding factor, with reference to two measurements of two samples of the same substance with the same atomic composition but different sizes and dimensions. In this aspect, the infinite dilution self-shielding factor at a value of $\bar{\ell}$ is obtained by reference to the difference $\bar{\ell}$ = $\bar{\ell}_{\text{large}}-\bar{\ell}_{\text{small}}$.

\begin{figure}[htb]
  \includegraphics[width=\linewidth]{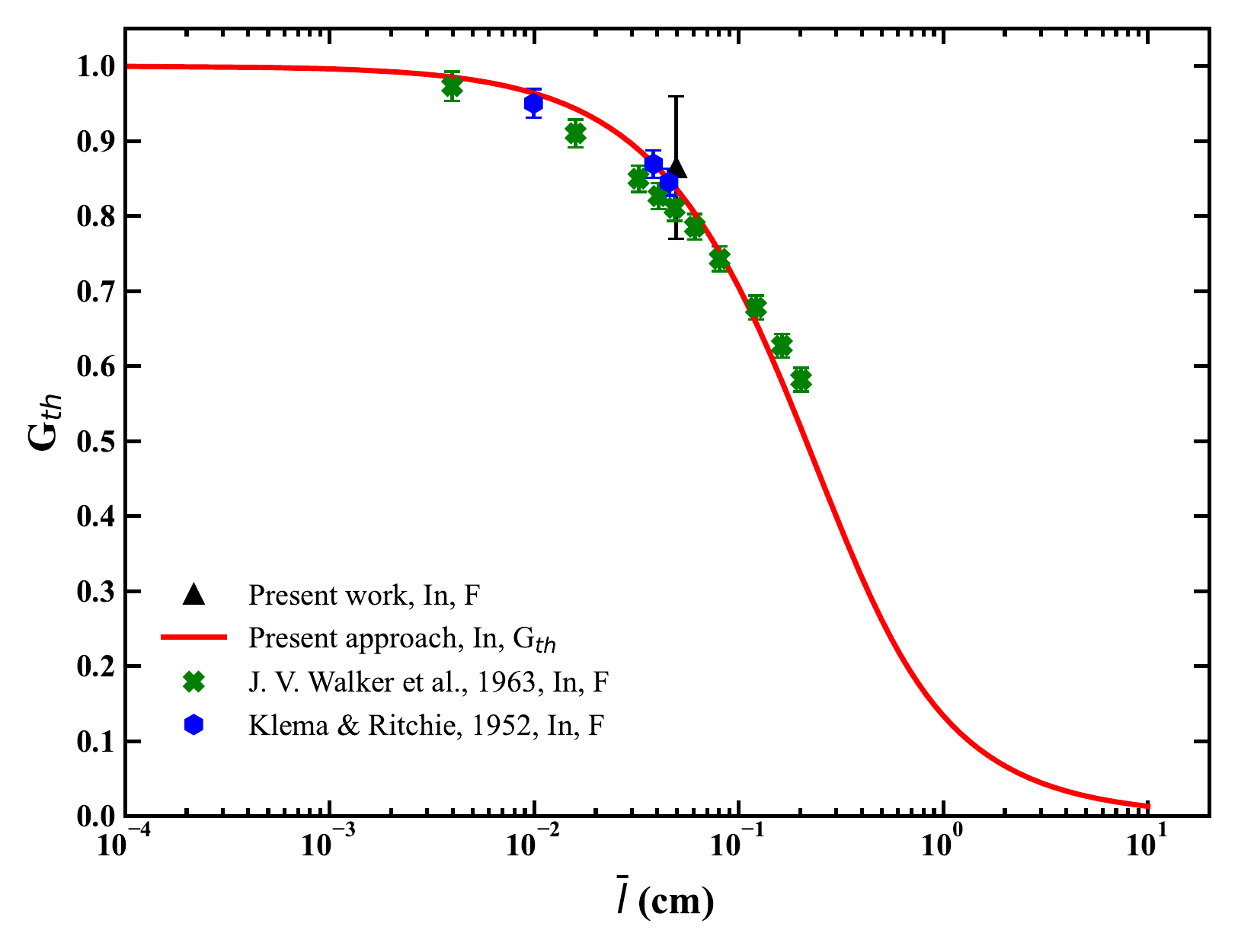}
  \caption{Comparison of G$_{th}$ of our approach and experimental measurement of indium foils with Experimental values taken from the literature. Experimental data of were those of Walker et al. \cite{walker1963thermal}, and  Klema \cite{klema1952thermal}. Error bars are either the digitization errors or a given uncertainty, see Section \ref{Sec:uncertainty}. (F: Foil).} \label{fig1}
\end{figure}

\begin{figure}[htb]
	\includegraphics[width=\linewidth]{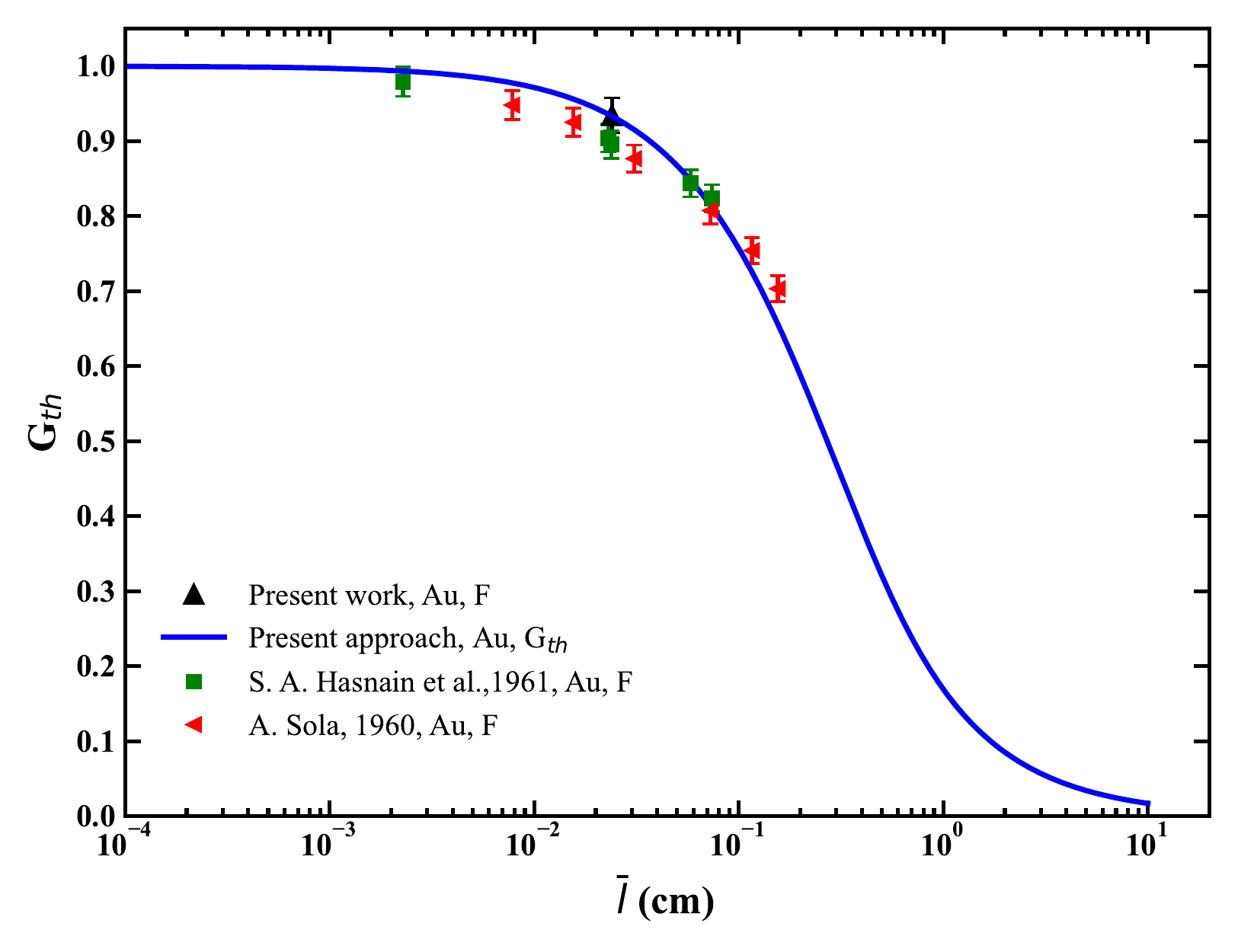}
	\caption{Comparison of G$_{th}$ of our approach and experimental measurement of gold foils with Experimental values taken from the literature. Experimental data of were those of Hasnain et al. \cite{hasnain1961thermal},and Sola \cite{Sola1960}. Error bars are either the digitization errors or a given uncertainty, see Section \ref{Sec:uncertainty}. (F: Foil).}\label{fig2}
\end{figure}

\begin{figure}[htb]
	\includegraphics[width=\linewidth]{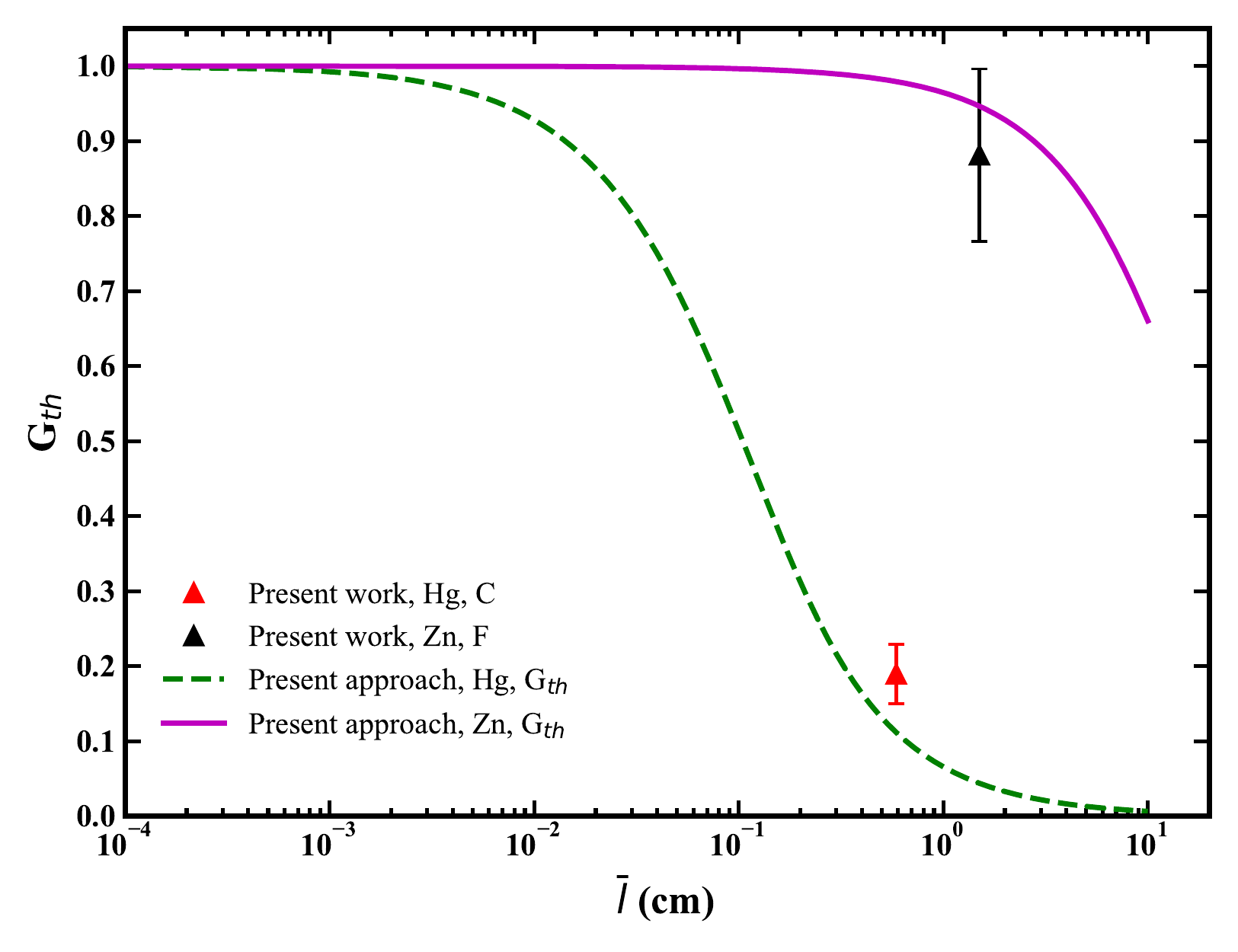}
	\caption{Comparison of G$_{th}$ of our approach and experimental measurement of zinc foils and  mercury cylinders. Error bars are a given uncertainty, see Section \ref{Sec:uncertainty}. (F: Foil and C: Cylinder).} \label{fig3}
\end{figure}

\begin{figure}[htb]
	\includegraphics[width=\linewidth]{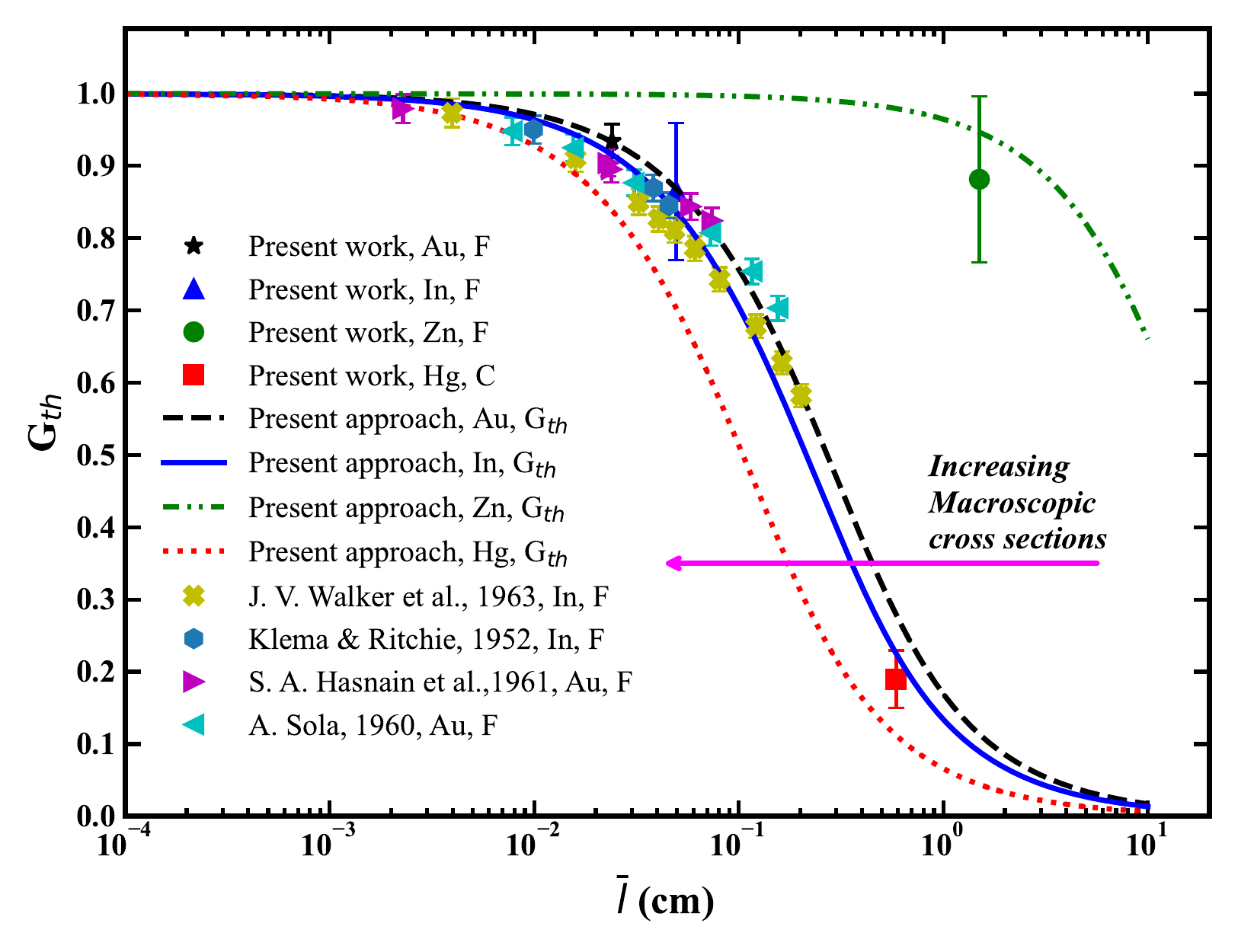}
	\caption{Comparison of G$_{th}$ of our approach and experimental measurement of indium, gold and zinc foils and  mercury cylinders with experimental values taken from the literature. Experimental data were those of Walker et al. \cite{walker1963thermal}, Klema \cite{klema1952thermal}, Hasnain et al. \cite{hasnain1961thermal},and Sola \cite{Sola1960}. The error bars are either the digitization errors or a given uncertainty, see Section \ref{Sec:uncertainty}. (F: Foil and C: Cylinder).} \label{fig4}
\end{figure}

Hence, the present experimental values of 0.865$\pm$0.095 for the indium self-shielding factor and 0.934$\pm$0.023 for the indium self-shielding factor are assigned to values of $\bar{\ell}$ = 0.0495 cm and 0.024 cm, respectively. Similarly, metallic zinc having $\bar{\ell}$ = 1.5013 cm corresponds to the self-shielding factor of 0.881$\pm$0.115,  while the value of $\bar{\ell}$ = 0.5885 cm of mercury corresponds to self-shielding factor of 0.1898$\pm$0.0396. These values of the neutron chord-length are used in the illustrations in Figs. \ref{fig1}-\ref{fig4}.

For the thermal neutron self-shielding factor in In, Au, Zn and Hg samples, experimentally measured and theoretically derived data were collected from various sources. Foils and cylinders were used as the geometry for these elements and the results are compared in Figs. \ref{fig1}, and \ref{fig2} for indium and gold, respectively, and in \ref{fig3} for zinc and mercury. The experimental data of G$_{th}$ of Hasnain et al. \cite{hasnain1961thermal}, Sola \cite{Sola1960},  Walker et al. \cite{walker1963thermal},and Klema \cite{klema1952thermal} for indium and gold foils were digitized from Martinho et al. \cite{Martinho2004637}.
There was no previously published experimental data for zinc and mercury self-shielding factors, see Fig. \ref{fig3}. The comparison between experimental results and the formula in Eqs. \ref{Eq:Selfshielding} and \ref{Eq:ApproximateP0}, in the thermal energy domain, gave an excellent agreement between the experimental data for In and  Au foils within the experimental uncertainty and the sigmoid function in Eq. \ref{Eq:Selfshielding} derived in Ref. \cite{MahmoudElmaghrabySalamaElghazalyElFikiarxiv220413239}. These isotopes had  accurate data for scattering and absorption cross-section together with the limited number of their isotopes (1 in the case of gold and 2 in the case of indium in which the most abundant is used as the monitored reaction), see Table \ref{ElementCrossSections}. Zinc and mercury, on the other hand, have numerous stable isotopes, 5 in case of zinc and 7 in case of mercury. Among these isotopes the monitored reaction focuses on one which has the most known value of cross-section and decay data, as illustrated in Table \ref{ElementCrossSections}. The reaction and decay data of other isotopes of zinc and mercury are affected by the purity of the isotope while conducting data evaluation. This explains the slight deviation from the self-shielding function as illustrated in Fig. \ref{fig3}.

A comparison between the variation of neutron self-shielding factor (sigmoid function in Eq. \ref{Eq:Selfshielding} and the interaction probability of Eq.  \ref{Eq:ApproximateP0}) for the four investigated elements together with their available experimental data  is given in Fig. \ref{fig4}. This emphasizes the importance of expressing the value of the thermal neutron self-shielding factor in terms of the average chord length of the absorbing body. Specifically speaking, the trend shows a sigmoid function that shifts to a higher average chord length, the $\bar{\ell}$, as the cross-section decreases. Or, on contrary, it shifts to shorter values  $\bar{\ell}$, as the cross-section increases because the interaction probability is used to correlate the neutron migration length in the convex-shaped absorber, it would be in inverse proportionality with the thermal neutron cross-section.

\section{Conclusion}
The agreement between present experimental results, data collected from literature, and the mathematical model of thermal neutron self-shielding factor suggests the validity of the mathematical model. The simplicity introduced by the use of average chord length in determining the value of neutron self-shielding based on integral parameter representation, which is available in nuclear data, makes the sigmoid function
  \[G_{\mathrm{th}} =\left(\frac{\Sigma_t}{\Sigma_s}\right)\times\left({1+\cfrac{\bar{\ell}}{(1-\exp(-\Sigma_t \bar{\ell}))}\cfrac{\Sigma_t\Sigma_a}{\Sigma_s}}\right)^{-1},\]
in thermal neutrons energy range, a fast and reliable technique in obtaining the self-shielding factor. The additivity property of the average chord length, which can be considered a measure of the migration length of neutrons inside the material, enabled correlating the measurements of infinite dilution self-shielding factor to samples having undiluted reference samples as directed from the agreement between values of $G_{th}$ for zinc and mercury and the expected values obtained mathematically. In samples with high macroscopic absorption cross-sections, the neutron self-shielding must be determined, especially in nuclear analytical techniques.

%

\end{document}